\documentclass[prl,superscriptaddress,showpacs,twocolumn]{revtex4}
\usepackage{graphicx}
\usepackage{graphicx,amsfonts}
\usepackage{epsfig,amsmath}
\usepackage{verbatim}

\begin{document}

\newcommand{\boldeps}{\boldsymbol\epsilon}
\newcommand{\atanh}
{\operatorname{atanh}}
\newcommand{\ArcTan}
{\operatorname{ArcTan}}
\newcommand{\ArcCoth}
{\operatorname{ArcCoth}}
\newcommand{\Erf}
{\operatorname{Erf}}
\newcommand{\Erfi}
{\operatorname{Erfi}}
\newcommand{\Ei}
{\operatorname{Ei}}

\title{Exact distribution of the maximal height of $p$ vicious walkers}  

\author{Gr{\'e}gory Schehr}
\affiliation{Laboratoire de Physique Th\'eorique (UMR du
  CNRS 8627), Universit\'e de Paris-Sud, 91405 Orsay Cedex,
  France}

\author{Satya N. Majumdar}
\affiliation{Laboratoire de Physique Th\'eorique et et Mod\`eles
  Statistiques, Universit\'e Paris-Sud, B\^at. 100, 91405 Orsay Cedex,
France}

\author{Alain Comtet}
\affiliation{Laboratoire de Physique Th\'eorique et et Mod\`eles
  Statistiques, Universit\'e Paris-Sud, B\^at. 100, 91405 Orsay Cedex,
France}

\author{Julien Randon-Furling}
\affiliation{Laboratoire de Physique Th\'eorique et et Mod\`eles
  Statistiques, Universit\'e Paris-Sud, B\^at. 100, 91405 Orsay Cedex,
France}

\date{\today}

\begin{abstract}
Using path integral techniques, we compute exactly the distribution
of the maximal height $H_p$ of $p$ nonintersecting Brownian
walkers over a unit time interval in one dimension, both for excursions ($p$-watermelons
with a wall) and bridges ($p$-watermelons without a wall), for all integer $p\ge 1$.  
For large $p$, we show that $\langle H_p \rangle \sim \sqrt{2p}$ (excursions) 
whereas $\langle H_p \rangle \sim \sqrt{p}$ (bridges). 
Our exact results prove that previous numerical experiments only
measured the pre-asymptotic behaviors and not the correct asymptotic ones.
In addition, our method establishes a physical connection
between vicious walkers and random matrix theory.        
\end{abstract}
\maketitle

{\it Introduction.} Since the pioneering work of de Gennes~\cite{pgdg}, 
followed up by Fisher~\cite{fisher}, the subject 
of vicious  
(non-intersecting) random walkers has attracted a lot of interest 
among physicists.
It has been studied in the context of
wetting and melting~\cite{fisher}, networks of
polymers~\cite{essam} and fibrous structures~\cite{pgdg}, persistence
properties in nonequilibrium systems \cite{bray_winkler} and stochastic growth
models \cite{richards, ferrari}. There also exist connections
between the vicious walker problem  
and the random matrix theory (RMT) \cite{johansson, katori_rmt,tracy_excursions},   
including for instance Dyson's Brownian motion \cite{dyson}. These connections
to RMT have rekindled recent interest in the vicious walker
problem and have led to new interesting questions.
However, 
despite extensive recent mathematical literature on the subject,   
the connections to RMT have so far been established using mostly  
combinatorial approaches. Given the non-intersection constraint in the vicious walker problem, it is 
natural to expect a free Fermion approach to make its connection to RMT
physically more explicit. The aim of this Letter is to present such an
approach which, in addition, allows us to derive a variety of new exact results in the
vicious walker problem.    

Physically, one dimensional vicious walkers play an important role 
in describing the elementary topological excitations in the $p \times 1$
commensurate adsorbed phases close to the commensurate-incommensurate (C-IC)
transition~\cite{huse}. In the commensurate phase the elementary excitations
are pairs of dislocations at a given distance with $p$ nonintersecting domain walls emerging
from one and terminating at the other. This is just a `watermelon' configuration
of $p$ nonintersecting Brownian bridges (see Fig. 1b). The sizes of such
defects and their fluctuations become important near the phase transition.
An important quantity that characterises the transverse fluctuations of
the defect is
the maximal height of the $p$ vicious walkers in a fixed time (here time signifies
the fixed longitudinal distance between the pair). 
Such extreme value questions have recently 
been studied extensively for a {\em single} Brownian bridge or
an excursion (with certain constraints) 
in the context of the maximal height of a fluctuating
interface~\cite{satya_airy,evs}.   
In this Letter, we obtain exactly the distribution of the maximal
height for $p$ nonintersecting  
Brownian bridges and excursions.

\begin{figure}
\includegraphics[width=\linewidth]{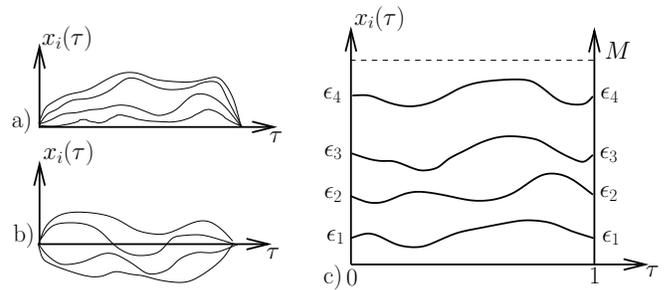}
\caption{a) : $4$-watermelons with a wall. b) :
  $4$-watermelons without wall. c) : illustration of the method to
  compute $F_4(M)$ using path integral techniques where appropriate
  cut-offs $\epsilon_i$'s have been introduced.}\label{fig1} 
\end{figure}

Motivated by the geometry of elementary excitations discussed above, we thus focus on 
``watermelons'' configurations (see Fig. \ref{fig1} a) and b)) where 
$p$ non-intersecting Brownian walkers $x_1(\tau)<  \cdots<
x_p(\tau)$, starting at $0$ at time $\tau =0$, arrive at the same position at $\tau~=~1$. We  
consider both
``$p$-watermelons with a wall'' (Fig. \ref{fig1} a)), where the
walkers stay positive in the time interval $[0,1]$ and
``$p$-watermelons without wall'' (Fig. \ref{fig1} b)) where the walkers are free to cross the origin
in between. 
Our main focus is on $H_p$, the maximal height 
of the top walker in $[0,1]$, $H_p = {\rm Max}_\tau [ x_p(\tau),\quad
  0 \leq \tau \leq 1 ]$. 

In particular, we are interested in the cumulative distribution
$F_p(M) = {\rm Proba.} \;[H_p \leq M]$ and in the moments $\langle 
H_p^s\rangle$. For $p=1$, there exist well known results\cite{borodin_book}, 
{\it e.g.} $\langle H_1\rangle = \sqrt{\pi/2}$ for an excursion, or $\langle H_1\rangle =
\sqrt{\pi/8}$ for a bridge. Recently, Bonichon and Mosbah (BM)
\cite{bonichon},  
using an algorithm based on exact enumerative formulas \cite{krattenthaler}, conjectured, from numerical simulations, that 
 for $p > 1$, $\langle H_p \rangle_{\rm num} \simeq
\sqrt{1.67p-0.06}$ for watermelons with a wall and  $\langle H_p \rangle_{\rm num} \simeq
\sqrt{0.82p-0.46}$ for watermelons without wall. These results stimulated several recent works 
\cite{fulmek, katori_p2, feierl1, feierl2, finch} aiming at an analytical derivation of these 
estimates.

On the other hand, exploiting the recent connection between watermelons and the Airy processes 
\cite{praehofer, 
tracy_excursions}, setting $\tilde x_p(\tau) = x_p(\tau)/\sqrt{\tau(1-\tau)}$, 
one expects that, in the limit $p \to \infty$, $\tilde x_p(\tau) =
A\sqrt{p} + p^{-1/6} \xi$ where $A=2^{3/2}$ (excursions) and $A=2$
(bridges), where $\xi$ is the Airy$_2$
process \cite{praehofer, tracy_excursions} of a suitably rescaled time
parameter. 
Thus in the large $p$ limit, the top curve approaches a limit shape, $x_p(\tau)\to A\sqrt{p} 
\sqrt{\tau(1-\tau)}$. Since the maximum of the top curve
occurs at the midpoint $\tau=1/2$, one expects that
for $p \gg 1$,
$\langle~H_p~\rangle~\sim~\langle~x_p(\tau~=~\tfrac{1}{2})\rangle~\sim~\sqrt{2p}$ for excursions and, similarly, $\langle H_p \rangle \sim
\sqrt{p}$ for bridges. These exact asymptotic estimates differ considerably 
from the numerical estimates of BM suggesting that the latter 
only describe the preasymptotic behavior
of $\langle H_p \rangle$. However, it calls for an explanation why this
preasymptotic  
behavior as measured by BM should be about
$\sqrt{1.67 p}$ and $\sqrt{0.82 p}$.   

In this Letter, we present a method based on path integrals associated to
corresponding free Fermions 
models to compute exactly $F_p(M)$. Our exact formula
is useful for a number of reasons. It provides the exact asymptotic tails
of the distribution of $H_p$ which were not known before. For the average height,
our formula
explains the aforementioned 
discrepancy between the estimates of BM and the exact asymptotic behaviors of $\langle H_p \rangle$. 
We show that for
moderate values of $p$ (preasymptotic behavior), one obtains $\langle H_p \rangle \propto
{\pi}\sqrt{p/6} = \sqrt{1.64493 \cdots p}$ for excursions 
and $\langle H_p \rangle \propto {\pi}\sqrt{p/12} = \sqrt{0.822467
  \cdots p}$ for bridges, in nice agreement 
with BM's estimates. 
Finally, we show how our method allows for a physical derivation of the
connection between 
$p$-watermelons configurations and RMT. 

{\it Method}. To calculate the cumulative distribution $F_p(M)$, we
 use a path integral method which needs to be 
 suitably adapted to this problem. Indeed one notices that the $p$-watermelons
 configurations described above (see {\it e.g.} Fig. \ref{fig1} a) and
 b)) are ill defined for systems in continuous space and time. For such
 Brownian walks, it is well known that if two walkers cross each other
 once, they will re-cross each other infinitely many times immediately
 after the first crossing. Therefore, it is
 impossible to enforce 
the constraint $x_{i}(0)=x_{i+1}(0)=0$ and simultaneously
 forcing $x_i(\tau) < x_{i+1}(\tau)$ immediately after. The
 cleanest way to circumvent this problem is to consider discrete time
 random walks 
 moving on a discrete one-dimensional lattice (so called Dyck
 path) : this is the method used in Ref. \cite{krattenthaler,
 fulmek, feierl1, feierl2}. By taking the 
 diffusion continuum limit, one would then arrive at non intersecting
 Brownian motions \cite{gillet}. This method is however mathematically cumbersome. Alternatively, following
 Ref. \cite{satya_airy, 
 satya_review}, we can go around this problem by
 assuming that the starting and finishing positions of the $p$ walkers
 are $0 < \epsilon_1  <  ... < \epsilon_p$ (see
 Fig. \ref{fig1} c)). Only at the end we take the limit $\epsilon_i
 \to 0$ and show that it is well defined. In addition, in
 order to compute $F_p(M)$, we
put an absorbing hard wall at $M$ such that
\begin{eqnarray}\label{starting_ratio}
F_p(M) = \lim_{\epsilon_i \to 0}
\left[\frac{N(\boldeps,M)}{N(\boldeps,M\to \infty)}
  \right] \;, 
\end{eqnarray}
where ${\boldeps} \equiv \epsilon_1, \cdots, \epsilon_p$ and 
$N(\boldeps,M)$ is the probability that the $p$
Brownian paths starting at $0 < \epsilon_1 <  ... <
\epsilon_p$ at $\tau =0$ come back to the same points at $\tau = 1$ without crossing each other and
staying within the interval $[\mu,M]$, with $\mu =0$ for excursions and $\mu
\to-\infty$ for bridges. This procedure is depicted in
Fig. \ref{fig1} c). 

The probability measure associated to $p$
unconstrained Brownian paths $x_1(\tau),.., x_p(\tau)$ over the time
interval $[0,1]$ is proportional to $\exp{[-\tfrac{1}{2}
    \sum_{i=1}^p \int_0^1 \left( \tfrac{dx_i}{d\tau}\right)^2 d\tau]}$. Here, we have
to incorporate the constraint that they stay in the interval
$[\mu,M]$. Therefore one can use path-integral techniques to write  
$N(\boldeps,M)$ in Eq. (\ref{starting_ratio}) as the propagator
\begin{eqnarray}\label{start_path_int}
N(\boldeps,M) = \langle \boldeps | e^{-\hat H_M}|\boldeps \rangle \;,
\end{eqnarray}
with $\hat H_M =  \sum_{i=1}^p [\tfrac{-1}{2}\tfrac{\partial^2}{\partial x_i^2} + V(x_i)]$, where $V(x)$ is a confining potential with $V(x) = 0$ if $x \in [\mu,M]$ and $V(x) = \infty$ outside this interval. Denoting by $E$ the eigenvalues of $\hat H_M$ and $|E\rangle$ the corresponding eigenvectors one has 
\begin{eqnarray}\label{path_int_formula}
N(\boldeps,M) = \sum_E |\Psi_E(\boldeps)|^2 e^{-E} \;,
\end{eqnarray}
where we introduced the notation $\langle \mathbf{x} | E \rangle =
\Psi_E(\mathbf{x})$. Importantly, to take into account the fact that we are
considering here non-intersecting Brownian paths, the many body wave
function $\Psi_E(\mathbf{x}) \equiv \Psi_E(x_1,..,x_p)$ must be
Fermionic, {\it i.e.} it 
vanishes if any of the two coordinates are equal. This many-body
antisymmetric wave function is thus constructed from the one-body
eigenfunctions of $\hat H_M$ by forming the associated {Slater determinant}.

{\it Watermelons with a wall.} In that case $\mu =0$ and the one-body
eigenfunctions are given by $\phi_n(x) =
\sqrt{\tfrac{2}{M}} \sin \tfrac{n \pi x}{M}$ with discrete eigenvalues
$\tfrac{n^2 \pi^2}{2 M^2}$, $n \in {\mathbb{N}^*}$. Therefore one has  
\begin{eqnarray}\label{slater_be}
\Psi_E(\boldeps) = \frac{1}{\sqrt{p !}} \det_{1 \leq i,j \leq p}
  \phi_{n_i}(\epsilon_j) \; , \; E = \frac{\pi^2}{2 M^2} {\mathbf n}^2  
\end{eqnarray}
where we use the notation $\mathbf{n}^2 = \sum_{i=1}^p n_i^2$, $n_i
\in \mathbb{N^*}$.   
From this expression (\ref{slater_be}), one checks that, in the limit $\epsilon_1 \to 0, \cdots, \epsilon_p \to 0$,  powers of $\epsilon_i$'s cancel
between the numerator and the denominator in
Eq. (\ref{starting_ratio}), yielding 
\begin{eqnarray}\label{expr_fp_be1}
&&F_p(M) = \frac{A_p}{M^{2p^2+p}} \sum_{n_1,\cdots,n_p} [\Xi({\mathbf
      n})]^2 e^{-\frac{\pi^2}{2M^2}{\mathbf{n}}^2} \;, \\ \nonumber 
&&\Xi({\mathbf n}) = \prod_{1\leq j<k \leq p} (n_j^2-n_k^2)
  \prod_{i=1}^p n_i \;,
\end{eqnarray}
where $A_p$, a constant independent of $M$, is determined by requiring
that $\lim_{M \to \infty} F(M) = 1$. It can be evaluated using a 
Selberg's integral \cite{mehta} yielding $A_p =
{\pi^{2p^2+p}}/[{2^{p^2-p/2}}{\prod_{j=0}^{p-1} \Gamma(2+j)
  \Gamma(\tfrac{3}{2}+j)}]$. 
For $p=1$, our expression gives back the well known result for a
Brownian excursion 
\cite{chung}. For $p=2$, we have 
checked, using the Poisson summation formula that our expressions in Eq.~(\ref{expr_fp_be1}) yield back the result of
Ref. \cite{katori_p2}. For generic $p$, the probability distribution function (pdf) $F'_p(M)$ is bell-shaped, exhibiting a single mode. 
At variance with previous studies \cite{katori_p2, feierl1}, 
our expression (\ref{expr_fp_be1}) is easily amenable to an asymptotic analysis for small $M$. Indeed, when $M \to 0$, the leading
contribution to the sum in (\ref{expr_fp_be1}) comes from $n_i =
i$ and its $p !$ permutations, yielding for $M \to 0$  
\begin{eqnarray}\label{asympt_fp_small}
&&F_p(M) \sim \frac{\alpha_p}{M^{2p^2+p}} e^{-\frac{\pi^2}{12 M^2} p
    (p+1) (2p+1)} \;, 
\end{eqnarray}
where $\alpha_p$ can be explicitly computed, yielding for instance 
$\alpha_2 = 12 \pi^9$. For large $M$, one can use the Poisson
summation formula to obtain $1-F_p(M) \propto  \exp(-2M^2)$. 

From the distribution in Eq. (\ref{expr_fp_be1}), one can compute the
moments of the distribution $\langle H_p^s \rangle$. For $p \geq 2$, one
obtains that $\langle H_p \rangle$ can be expressed in terms of
integrals involving 
the Jacobi theta function $\vartheta(u) = \sum_{n=-\infty}^\infty e^{-\pi
  n^2 u}$ and its derivatives, thus recovering, by a
simpler physical derivation, the results of 
Ref. \cite{fulmek, katori_p2, feierl1, feierl2}. In particular, one has $\langle H_2 \rangle =
1.82262...$ \cite{fulmek}.
For moderate values of $p$, one observes that the main contribution to the average $\langle H_p
\rangle = \int_0^\infty M F'_p(M) dM$ comes from relatively small $M$ where 
$F'_p(M)$ is dominated, as before in Eq.~(\ref{asympt_fp_small}), by the
terms where $n_i = i$ and its $p!$ permutations. It is easy to see that the pdf, restricted to this first term (\ref{asympt_fp_small}) exhibits a 
maximum for $M^* \sim \pi\sqrt{p/6}$. Therefore, one expects that $\langle H_p \rangle \sim M^*  =
\sqrt{1.64493 \cdots p}$, in good 
agreement with the estimates of BM \cite{bonichon}. For larger values
of $p$ the average $\langle H_p \rangle$ picks up contributions from
larger values of $M$ where $F'_p(M)$ can not be approximated by a
single term as in Eq.~(\ref{asympt_fp_small}) and therefore the estimate
of BM ceases to be correct. Instead, one has the exact asymptotic
behavior $\langle H_p \rangle \sim \sqrt{2p}$ for $p \gg 1$, which can be
obtained directly from our formula in Eq. (\ref{expr_fp_be1})
\footnote{Details will be published elsewhere.}.  

{\it Watermelons without wall.} In the case of Brownian bridges, one
can apply the same formalism as above
(\ref{starting_ratio})~-~(\ref{path_int_formula}) with
$\mu~\to~-\infty$, {\it i.e.} $\hat H_M = \sum_{i=1}^p
\tfrac{-1}{2}\tfrac{\partial^2}{\partial 
  x_i^2}$. In that case, the 
one-body eigenfunctions are given by $\psi_k(x) =
\sqrt{\tfrac{2}{\pi}} \sin{[k (M-x)]}$ with a continuous spectrum
$E_k = k^2/2$, $k \in {\mathbb{R^+}}$. Therefore, $\Psi_E(\boldeps)$
entering the expression of 
$N(\boldeps, M)$ in Eq.~(\ref{path_int_formula}) is
formally given by Eq.~(\ref{slater_be}) where $\phi_{n_i}$
is replaced by $\psi_{k_i}$ and $E =
\tfrac{{\mathbf k}^2}{2}$. One
obtains 
\begin{eqnarray}\label{expr_fp_bb} 
&&F_p(M) = \frac{B_p}{M^{p^2}} \int_0^\infty \,dy_1 \cdots
  \int_0^\infty \,dy_p  \, e^{-\tfrac{{\mathbf{y}^2}}{2 M^2}}
  \Theta_p({\mathbf{y}})^2 \;, \nonumber \\
&&\Theta_p({\mathbf{y}}) = \det_{1 \leq i,j \leq p} y_i^ {j-1} \cos(y_i + j \tfrac{\pi}{2} ) \;,
\end{eqnarray}
where $B_p = 2^{2p}/[(2\pi)^{p/2} \prod_{j=1}^p \Gamma(j+1)]$. This
yields, for instance, $F_2(M) = 1 - 4M^2e^{-2M^2} - e^{-4
  M^2}$. From Eq. (\ref{expr_fp_bb}), one obtains the asymptotic behavior for $M \to 0$ as 
\begin{eqnarray}\label{small_bb}
F_p(M) \propto M^{p^2+p} \;,
\end{eqnarray}
whereas for large $M$ one has $1-F_p(M) \propto \exp{(- 2 M^2)}$. As in
the case of watermelons with a wall, the pdf $F_p'(M)$ is also 
bell-shaped with a single mode. Notice however that the presence of the
wall has drastic effects on the small $M$ behavior of $F_p(M)$ (see
Eq.~(\ref{asympt_fp_small}) and Eq.~(\ref{small_bb})) whereas, as
expected, it has
less influence for large $M$.   
\begin{figure}[h]
\includegraphics[width=\linewidth]{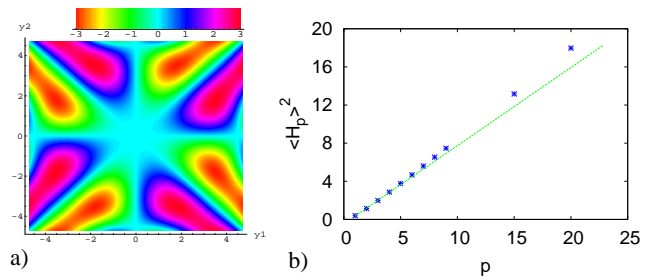}
\caption{a) : Contour plot of $\Theta_2(y_1,y_2) = y_2 \sin{y1}\cos{y2}-y_1
  \cos{y_1} \sin{y_2}$ given in (\ref{expr_fp_bb}). It exhibits saddles 
  for $(y_1,y_2) = (\pm \pi/2,\pm \pi)$ and symmetric points obtained
  by permutations. b) Plot of $\langle H_p \rangle^2$ as a function of $p$. The dotted line is the
estimate from BM \cite{bonichon}. The quality of this estimate for $p
  \lesssim 10$ has its origin in the saddles
  of $\Theta_p({\mathbf{y}})$ shown, for $p=2$,
  on the left panel. For larger values of $p$ one has instead $\langle
  H_p \rangle^2 \propto p$.}\label{fig3}
\end{figure}

From $F_p(M)$ in~(\ref{expr_fp_bb}), one computes the moments
$\langle H_p^s \rangle$, yielding 
$\langle H_2 \rangle = \tfrac{1+\sqrt{2}}{4} \sqrt{\pi}$
or $\langle H_3 \rangle = \tfrac{45+36\sqrt{2} - 8\sqrt{6}}{96}
\sqrt{\pi}$, recovering (to leading order) recent results obtained by rather
involved 
combinatorial techniques \cite{feierl2}.  

To make contact with BM's estimates, one first focuses on $p=2$ and 
notices that $\Theta_2(y_1,y_2)$ in Eq. (\ref{expr_fp_bb}) exhibits saddles
for $y_1 = \pm \pi/2, y_2 
= \pm \pi$ and for symmetric points obtained by permutations: this is shown in
Fig. \ref{fig3} a). In fact this
property can be generalized to higher values of $p$ and one
can show that $\Theta_p({\mathbf{y}})$ has saddles which are located
around $y_1 = 
\pm \pi/2, y_2 = \pm \pi, \cdots , y_p = \pm p \pi/2$ and the points
obtained by permutations. Of course $\Theta_p(\mathbf y)$ develops saddles for
higher values of ${\mathbf y}^2$ but their weights are exponentially
suppressed in Eq. (\ref{expr_fp_bb}). For moderate values of $p$, one
expects that $\langle H_p \rangle$ is dominated by these saddles $y_1 = 
\pm \pi/2, y_2 = \pm \pi, \cdots , y_p = \pm p \pi/2$. Therefore
performing a saddle point calculation, one has $
F_p(M) \propto e^{-p^2 \chi\left(\frac{M}{\sqrt{p}}\right)}$, with  
$\chi(y) = \log{y} + {\pi^2}/{(24 y^2)}$, which has a minimum for $y^* =
\pi/\sqrt{12}$. This yields $\langle H_p \rangle \sim {\pi}\sqrt{p/12} =
\sqrt{0.822467 \cdots p}$, in good agreement with the estimates of
BM \cite{bonichon}. For larger values of $p$ one expects
that $\langle H_p \rangle$ picks up contributions from larger values
of $M$ where $F_p(M)$ can not be reduced to these first saddles. In
Fig. \ref{fig3} b), one shows a comparison 
between the exact value of $\langle H_p \rangle^2$ computed from
Eq. (\ref{expr_fp_bb}) and the estimate of BM. 
This clearly shows that the estimate of BM correspond to the pre-asymptotic behavior. Instead, for large $p$, one expects here
$\langle H_p \rangle \propto \sqrt{p}$.

{\it Extension of the method}. The method presented here can be used 
to derive many other results. As an
interesting example, showing explicitly 
the connection between watermelons and RMT, we compute the joint
probability 
distribution $P_{\rm joint}(x_1, 
\cdots, x_p,\tau)$, first for $p$ bridges. Following the same
steps as above, Eq. (\ref{starting_ratio})-(\ref{path_int_formula}), and using
the Markov property of 
Brownian paths, one has 
\begin{eqnarray}
P_{\rm joint}({\mathbf x},\tau) = \lim_{\epsilon_i \to 0}
  \frac{\langle \boldeps |e^{-\tau \hat H_0}| {\mathbf x} \rangle
  \langle {\mathbf x} | e^{-(1-\tau) \hat H_0} | \boldeps \rangle   }{
\langle \boldeps |e^{-\hat H_0}| \boldeps \rangle} \label{starting_joint}
\end{eqnarray} 
with $\hat H_0 = \sum_{i=1}^p \tfrac{-1}{2}\tfrac{\partial^2}{\partial
  x_i^2}$. One can show that powers of $\epsilon_i$'s
  cancel between the numerator and the denominator in (\ref{starting_joint}), yielding $P_{\rm
  joint}({\mathbf x},\tau) \propto Q({\mathbf x},\tau)  Q({\mathbf
  x},1-\tau) $ with
\begin{eqnarray}\label{expr_q}
Q({\mathbf x},\tau) = \int d \mathbf k \prod_{i<j} (k_i-k_j)
  e^{-\tfrac{\tau {\mathbf k}^2}{2} }  \det_{1\leq m,n \leq p }
  e^{(i x_m k_n )},
\end{eqnarray} 
where $\int d \mathbf k \equiv \int _{-\infty}^\infty dk_1 \cdots \int
_{-\infty}^\infty dk_p$. After some algebra to evaluate the integrals
in Eq. (\ref{expr_q}) one 
finally obtains, for $p$-watermelons without wall 
\begin{equation}\label{final_gue}
P_{\rm joint}({\mathbf x},\tau) = Z_p^{-1} \sigma(\tau)^{-p^2}\prod_{i<j}
(x_i-x_j)^2 
e^{-\frac{{\mathbf x}^2}{2 \sigma^2(\tau)}} \;,
\end{equation}
with $\sigma(\tau) = \sqrt{\tau(1-\tau)}$ and $Z_p$ a normalization
constant. This expression in Eq. (\ref{final_gue}) shows
that this joint probability is exactly the one of the eigenvalues
of the Gaussian Unitary Ensemble of random matrices (GUE) \cite{johansson,
  katori_rmt, dyson}. In particular, for $p \gg 1$, defining the
rescaled variable $\eta =
\sqrt{2}p^{1/6}(\tfrac{x_p(\tau)}{\sqrt{2}\sigma(\tau)}- \sqrt{2p})$,
one obtains that the
cumulative distribution of $\eta$ is given by ${\rm Proba}[\eta \leq x] = {\cal
  F}_{2}(x)$, the Tracy-Widom distribution for
$\beta=2$ \cite{tracy_widom}.  

For excursions, a similar calculation shows that
\begin{equation}\label{final_wishart}
P_{\rm joint}({\mathbf x},\tau) = {Z}'^{-1}_p \sigma(\tau)^{-p(2p+1)}
[\Xi({\mathbf x})]^2 e^{-\frac{{\mathbf x}^2}{2 \sigma^2(\tau)}} \;,
\end{equation}
where $\Xi({\mathbf x})$ is defined in (\ref{expr_fp_be1}) and
$Z'_p$ a normalization constant. Hence the joint distribution of $y_i =
x_i^2/2 \sigma^2(\tau)$ is formally identical to the distribution of the
eigenvalues of Wishart matrices \cite{mehta} with $M-N = \tfrac{1}{2}$, and $N=p$. In
that case, from the results for the largest
eigenvalue of Wishart matrices we conclude that for $p\gg1$, the
cumulative distribution of the rescaled variable $\zeta =
2^{2/3}p^{1/6}(\tfrac{x_p(\tau)}{\sqrt{2}\sigma(\tau)}- 2\sqrt{p})$
is again given by ${\cal F}_2(x)$~\cite{max_wishart}.

{\it Conclusion.} To conclude, using methods of many-body physics,
where appropriate cut-offs $\epsilon_i$'s have been introduced (see
Fig. \ref{fig1} c)),  
we have obtained exact results for the distribution of the maximal
height for $p$-watermelons with a wall (\ref{expr_fp_be1}) and
without wall (\ref{expr_fp_bb}), which is physically relevant to describe the
geometrical 
properties of dislocations arising in $p \times 1$ commensurate adsorbed phases
close to the C-IC transition. Our expressions explain the
discrepancy between the estimates of BM \cite{bonichon} and the true
asymptotic behaviors for the average $\langle H_p \rangle$. Besides,
we obtained a quantitative description of the pre-asymptotic regime
actually measured in the numerical experiments of BM. We hope that the path
integral method presented here, which is rather general, 
and the precise connection to RMT 
will allow further future~studies. 
\acknowledgements
We thank P.~Ferrari for useful discussions.

\end{document}